# The structural, magnetic and optical properties of $TM_n@(ZnO)_{42}$ (TM = Fe, Co and Ni) hetero-nanostructure

Yaowen Hu[a], Chuting Ji[a], Xiaoxu Wang[b, c, †], Jinrong Huo[b, †], Qing Liu[b], and Yipu Song[d,*]

[a]Department of Physics, Tsinghua University, Beijing 100084, China

[b]Department of Physics, University of Science and Technology Beijing, Beijing 100083, China.

[c]Department of Cloud Platform, Beijing Computing Center, Beijing 100094, China

[d]Center for Quantum Information, IIIS, Tsinghua University, Beijing 100084, China

**Abstract:**

The magnetic transition-metal (TM) @ oxide nanoparticles have been of great interest due to their wide range of applications, from medical sensors in magnetic resonance imaging to photo-catalysis. Although several studies on small clusters of TM@oxide have been reported, the understanding of the physical electronic properties of $TM_n@(ZnO)_{42}$ is far from sufficient. In this work, the electronic, magnetic and optical properties of $TM_n@(ZnO)_{42}$ (TM = Fe, Co and Ni) hetero-nanostructure are investigated using the density functional theory (DFT). It has been found that the core-shell nanostructure $Fe_{13}@(ZnO)_{42}$, $Co_{15}@(ZnO)_{42}$ and $Ni_{15}@(ZnO)_{42}$ are the most stable structures. Moreover, it is also predicted that the variation of the magnetic moment and magnetism of Fe, Co and Ni in $TM_n@ZnO_{42}$ hetero-nanostructure mainly stems from effective hybridization between core TM-3d orbitals and shell O-2p orbitals, and a magnetic moment inversion for $Fe_{15}@(ZnO)_{42}$ is investigated. Finally, optical properties studied by calculations show a red shift phenomenon in the absorption spectrum compared with the case of $(ZnO)_{48}$.

**Keywords:** Core-shell structure; Magnetic properties; Electronic structure

------------------------------------------------------------

\* Corresponding author: *E-mail*: ypsong@mail.tsinghua.edu.cn
† These authors contribute equally to this work

# 1. Introduction

Semiconducting hybrid materials with improved functionalities such as optical, electric and magnetic properties have been considered as potential candidates for a wide range of applications. For example, Rh, Pd and Pt particles supported on oxides, such as $CeO_2$ and $Al_2O_3$, are widely used in catalysis [1,2]; magnetic iron-oxide nanoparticles have been investigated as contrast agents for magnetic resonance imaging [3], which is of important use in cancer therapy. In particular, the physical properties of ZnO doped with ions of transition metal elements have been one of the most intriguing research topics in current materials science [4-9]. The characteristics of ZnO with Zn being a transition metal enables it to easily dope magnetic transition metal (TM) ions such as $Mn^{2+}$, $Fe^{3+}$, $Co^{2+}$ and $Ni^{2+}$ in place of $Zn^{2+}$ in the crystal of ZnO. Dietl et al. discovered room temperature ferromagnetism in Mn- doped ZnO thin film, receiving tremendous attention to ZnO based materials [10]. Since then several studies have been carried out in ZnO based materials with different combinations of TM ions [4, 11, 12]. Some reports revealed the importance of point defects such as oxygen and zinc vacancies and interstitials in magnetic ordering [13, 14]. Mishra and Das [15] studied the optical characteristics of Fe-doped ZnO nanoparticles using FTIR. Sawalha et al. [16] investigated the electrical conductivity of pure and doped ZnO ceramic systems. Their experiment indicated that donor concentration, point defects, and adsorption–desertion of oxygen were affected by the Fe doping for ZnO. Moreover, Shi and Duan [17] studied the magnetic properties of TM (Cr, Fe and Ni) doped in ZnO nanowires by first-principles theory. Xiao et al. [18] calculated the structural and electronic properties of Fe-doped ZnO nanoparticles, and the results showed that Fe doped ZnO nanoparticles were structurally more stable than the isolated FeO and ZnO phases.

In recent years, core-shell structures in which metals form the core and ZnO constitutes the shell have attracted intense interest due to their significantly high effectiveness in improving the photo-catalytic activity and the synergistic effect among components [19-22]. The core-shell architecture avoids exposing the inner core to the environment and thus maximizes the interaction between the building blocks. Moreover, the composition, size and morphology of the inner core and outer shell are important aspects of structural property and would most probably affect its stability. So far, to the best of our knowledge, investigations on the physical mechanism for the effect of composition, size and morphology of magnetic TM-core/ZnO-shell heterogeneous nanoparticles are very rare. Here, we report the theoretical studies on a series of $TM_n@(ZnO)_{42}$ (TM = Fe, Co and Ni) heterostructures by using the density functional theory (DFT). The structural, magnetic and optical properties of such core-shell heterostructures have been investigated. Stable structures are founded among different models and variation of magnetic moment are studied, especially for the moment inversion of $Fe_{15}@(ZnO)_{42}$. Furthermore, a red shift phenomenon is also obtained for the absorption spectrum of $Fe_{15}@(ZnO)_{42}$ compared with the case of $(ZnO)_{48}$. We expect that our results for $TM_n@(ZnO)_{42}$ can help to understand the effects of the encapsulation on the structure, stability, and magnetic properties of TM clusters.

## 2. Results and Discussion

### 2.1 The structural properties of TM$_n$@(ZnO)$_{42}$ hetero-nanostructure

In simulation, due to the multiplicity and indeterminacy of core-shell hetero-structure, it is always a challenge to optimize the stable structure of metal-oxide heterogeneous with increasing number of atoms. In the following calculations, the TM$_n$@(ZnO)$_{42}$ core-shell model is built to investigate stable structure of TM$_n$@(ZnO)$_{42}$ with different n (n=6-18). Considering the rationality of the structure, the magic number nanostructure of (ZnO)$_{48}$ with $D_{3d}$ symmetry is firstly chosen to be the initial configurations due to the fact that the (ZnO)$_{48}$ models has the highest binding energy [29]. Therefore, six ZnO in the center of relaxed (ZnO)$_{48}$ are removed, and magnetic TM-core TM$_n$ clusters are constructed. The central empty position to put the magnetic TM-core relies on some physical and chemical intuition based on the symmetry of bond-length and structure. Then, these atomic geometries are fully optimized until the convergence criteria are reached. According to our scheme, the stable configurations of TM$_n$@(ZnO)$_{42}$ clusters are obtained as shown in Fig. 1. The inner core TM$_n$ and outer shell (ZnO)$_{42}$ configurations separated from the optimized geometry configurations of TM$_n$@(ZnO)$_{42}$ are also illustrated in Fig. 1. It is noted that the encapsulated TM$_n$ (n = 6-7 for Fe, n = 6−9 for Co and Ni) clusters shift towards the (ZnO)$_{42}$ inside surface, indicating the presence of an attractive interaction of the TM$_n$ clusters caused by the (ZnO)$_{42}$ inside surface. However, large TM$_n$ clusters (n = 8-16 for Fe, n = 10-18 for Co and Ni) are nearly located at the center of the cages due to the inner core TM$_n$ cluster and outer shell cage sizes. The shells of n from 6 to 12 are a cage-like structure while the shells of n≥13 has a tendency to change into a sphere, which may imply that with the increase of n, the shell is increasingly inclined to become a spherical structure. The exact symmetry for each TM cluster is C1 except that $Ni_{12}$ is C2. Furthermore, it is intriguing that, in the $TM_n$ clusters, the TM atom located at the prominent position and the center of $TM_n$ (yellow balls in Fig. 1) have relatively small local magnetic moments. Therefore, there is a strong tendency of the magnetic $TM_n$ clusters for lower symmetry structures, which helps to increase their energy stability due to the splitting of the highest occupied states. From the results of bond lengths (see Fig. 1), the $Fe_n$ clusters are much more non-compact than the $Co_n$ and $Ni_n$ structures, indicating that the core is more close to shell for $Fe_n@(ZnO)_{42}$. This trend may affect the magnetic moments (see Supporting Information II) of the $TM_n@(ZnO)_{42}$ systems and induce more abnormal effect. (e.g. the atom with larger local magnetic moments for Fe shows inversion for the Fe atoms close to O atoms).

To investigate the structural stability, second-order differences of total energies ($\Delta_2 E$) for $TM_n@(ZnO)_{42}$ nanostructure are calculated and displayed in Fig. 2. The second-order differences of total energies are calculated by equation (1):

$$\Delta_2 E_n = E_{n-1} + E_{n+1} - 2E_n \qquad (1)$$

where $E_n$ and $n$ refer to the total energy of $TM_n@(ZnO)_{42}$ and the number of TM atoms, respectively.

As shown in Fig. 2, the relatively large peaks of $\Delta_2 E_n$ are found at n= 13, 15 and 15 for TM = Fe, Co and Ni in $TM_n@(ZnO)_{42}$ core-shell structures, respectively, demonstrating that $Fe_{13}@(ZnO)_{42}$, $Co_{15}@(ZnO)_{42}$ and $Ni_{15}@(ZnO)_{42}$ are the most stable configurations among all the clusters in the size range of the present study. The calculated Zn-O bond lengths and O-Zn-O bond angle of the $(ZnO)_{48}$ and M@ZnO (we use M@ZnO to represent all $TM_n@(ZnO)_{42}$ for TM= Fe, Co, Ni and n = 13,15,15 in the following discussion) are listed in Table 1 together with other calculated work [29], from which it can be seen that our results of $(ZnO)_{48}$ reach an agreement with the other studies [29]. It is also obvious that there is a contraction behavior for the outer-shell of M@ZnO compared with $(ZnO)_{48}$, indicating that doping at the center with a magnetic TM atom could provide strong bonding among surface atoms, that is, the Zn-O bonding of M@ZnO is stronger than the $(ZnO)_{48}$ cluster due to the interaction of M-O.

## 2.2 The magnetic and electronic structure properties of $TM_n@(ZnO)_{42}$ hetero-nanostructure

The magnetic properties of encapsulated $TM_n$ (TM = Fe, Co and Ni) clusters inside $(ZnO)_{42}$ are calculated based on the stable geometries discussed above. All of the transition metal atom magnetic moments of the $TM_n@(ZnO)_{42}$ core-shell nanostructure are shown in Table 2 and 3. More details of magnetic moments are described in the Supporting Information II. The following trends can be observed: (i) Except for a few cases, the magnetic moments decrease from outside to inside for core transition metal atoms. For example, for relatively stable structure $Fe_{15}@(ZnO)_{42}$, $Co_{15}@(ZnO)_{42}$, and $Ni_{13}@(ZnO)_{42}$, the center Fe, Co, Ni atoms have the magnetic moments 1.996, 1.167, and 0.226μB/atom, which are significantly smaller than the other magnetic moments such as 2.64, 1.78, and 0.68μB/atom, the average value for Fe, Co and Ni, respectively. (ii) As is presented in Table 4, a general feature is that local magnetic moments tend to have some relationship with the TM-O distance and the small distance corresponds to a small magnetic moment. Especially for several $Fe_n@(ZnO)_{42}$ systems, e.g., $Fe_{15}@(ZnO)_{42}$, it is found that some Fe local magnetic solutions change from ferromagnetic to antiferromagnetic phases (e.g. -2.176μB /atom) with the Fe-O distance decreased. A similar phenomenon can also be found in the $TM@Mg_{12}O_{12}$ [30] and $TM_m@C_n$ [31]. (iii) For most of the systems, we observed a large number of atomic configurations with slightly different magnetic moments. These results indicate that one of the magnetic configurations might be more favorable or a wide range of magnetic configurations might exist at real experimental conditions, and experimental techniques might access only the average results.

To obtain a better understanding for the origin of TM magnetic moments difference, we take relatively stable compound mentioned above as examples to present the charge density difference and investigate the p-O and d-TM projected DOS (see Fig. 3 and Fig. 4). Charge transfer data of the typical atom has been marked out in Fig. 3, demonstrating that, apart from the transition metal atoms neighboring oxygen atoms, the charge transfer numbers increase from outside to inside for core transition metal atoms. For instance, the Fe, Co, Ni atoms at the center of core have the charge transfer numbers 0.1431, 0.2181 and 0.2155, which are much larger than the numbers of other transition metal atoms and correspond to smaller magnetic moments as discussed previously. More intriguingly, because of the interaction with O atoms, transition metal atoms near O atoms have a large charge transfer, leading to smaller magnetic moment and even magnetization reversal has been found.

Fig. 4 shows the PDOS of the representative atoms for up (↑) and down (↓) spins, demonstrating differences in the shape of PDOS among transition metal atoms at different position. This shape differences are mainly a shift to high energy or low energy, which can be explained by the decrease or increase of the effective hybridization between core TM-3d orbitals and the shell O-2p orbitals, resulting in the charge transfer from core TM to shell O. The charge density difference is demonstrated in Fig. 3. Near to the Fermi level ($E_F$), a large overlap between O 2p and TM 3d is clearly seen for atoms with minimal TM-O distance, showing a strong hybridization between O and TM atoms. This also explains why all the core-shell clusters have relative large core-shell interaction energy.

In the case of $Fe_{15}@(ZnO)_{42}$, as the interaction of TM-O increases, spin-split becomes significant for the d- projected DOS, resulting in a large magnetic moment change from the center of the core to the edge of the core. In Fig. 3, we take Fe5, Fe9 and Fe14 as an example. Fe5 is in the center of the core with a magnetic moment 1.996μB/atom. Fe9 and Fe14 are located on the edge of the core with a moment -2.714μB/atom and -0.767μB/atom, indicating an inverse direction compared with other moments. It can be seen that the O atoms neighboring Fe9 and Fe14 are O33 and O31 respectively. At the same time, these two O atoms have the largest two negative magnetic moments (-0.108μB/atom and -0.064μB/atom) and they have a very large charge transfer at the same time. So the moment inversion could be interpreted by the strong interaction between Fe-O atoms, which could also be utilized to understand the magnetic moment difference for different core atoms. Similar spin-split can also be seen in $Co_{15}@(ZnO)_{42}$ and $Ni_{13}@(ZnO)_{42}$, where the spin-up and spin-down DOS become asymmetric. For Co and Ni at more outer positions or with smaller TM-O distances, this asymmetry is more significant. For $Co_{15}@(ZnO)_{42}$ and $Ni_{13}@(ZnO)_{42}$, spin-split becomes less obvious because of the weaker TM-O interaction than $Fe_{15}@(ZnO)_{42}$, resulting in the less magnetic moment change. For instance, magnetic moment for $Co_{15}@(ZnO)_{42}$ is 1.167μB/atom for Co8 at the center of core, 1.861μB/atom (Co1) for atom at the edge of core but far from O atom. And 1.700μB/atom (Co7) for atom at the edge is close to O (O17) atom. At the same time, see Fig.3 and Fig.4, the moment of O (O17) close to Co (Co7) is 0.059μB/atom and the moment of O (O29) near Ni (Ni6) is 0.027μB/atom, which is not large enough and leads to less TM moment changes. Moreover, in Fig. 3, less charge transfer from core Co or Ni to shell O also shows weaker interaction than $Fe_{15}@(ZnO)_{42}$.

In addition, it is indicated from Table 4 and Supporting Information II that the magnetic moment of transition metal is related to the coordination number, the average TM-TM bond length and the distance of TM-O, e.g. large coordination number usually lead to small magnetic moment; and small bond length of TM-TM or TM-O also tend to result in a small moment. The variation moments of TM atom may arise from the contribution of the synergistic effect of the coordination number and bond length. For Co, Ni atoms at the center of the core, their moments are small due to the largest coordination number among all the core atoms. Although Co1 has a larger coordination number compared with Co11, the average bond length between Co1 and neighbor Co atoms (2.458Å) is larger than the case of Co11 with adjacent Co atoms (2.448Å). At the same time, the distance of Co-O atoms is largest for Co1-O. As a result, the final moment of Co1 is strongest in $Co_{15}@(ZnO)_{42}$, indicating that the magnetic moment of atoms are deeply related to the geometry configuration of each atoms, which is consistent with the result from charge transfer.

## 2.3 The optical properties of M@ZnO and $(ZnO)_{48}$

In order to investigate the influence of magnetic TM inner-core on the optical properties of ZnO shell cage, the dielectric function of M@ZnO and pure $(ZnO)_{48}$ nanostructures are all calculated for comparison and the optical absorption of core-shell structure and pure $(ZnO)_{48}$ is illustrated in Fig. 5. Compared with the $(ZnO)_{48}$, the optical absorption peaks of core-shell structure $Fe_{13}@(ZnO)_{42}$, $Co_{15}@(ZnO)_{42}$ and $Ni_{15}@(ZnO)_{42}$ have an obvious red shift at 147.56 nm, compared with pure $(ZnO)_{48}$ at 123.95 nm, which is due to the effect of Fe, Co, Ni core. The spectral line of the $Ni_{15}@(ZnO)_{42}$ appeared to have a smaller peak at about 255.84 nm while both $Fe_{13}@(ZnO)_{42}$ and $Co_{15}@(ZnO)_{42}$ have no peak there. According to the DOS of $Ni_{15}@(ZnO)_{42}$ (shown in Fig. 4(d)), we found that this smaller peak originates from the stronger interaction between Ni-O atoms and more abundant charge transfer of O – Zn atom (see Supporting Information II). We contrast the DOS of M@ZnO (Fig. 4(d)) and conclude that the influence on Zn-O interaction for the case of introducing Ni atoms is weaker than the case of Co, Fe atoms, particularly at < 3 eV. Moreover, it is noted that, at around the 400 nm (in the visible light), all the core-shell structures and $(ZnO)_{48}$ have a distinct peak, although with some differences in the height of the peak, coming from the contribution of electrons transfer of O – Zn atom at shell.

Fig.5 also exhibits the imaginary part and real part of dielectric function of M@ZnO and pure $(ZnO)_{48}$. For the real part of dielectric function of $Fe_{13}@(ZnO)_{42}$, $Co_{15}@(ZnO)_{42}$ and $Ni_{15}@(ZnO)_{42}$, It is found that there are no obvious differences among them. Moreover, both of the real parts of dielectric functions of $(ZnO)_{48}$ and M@ZnO are all positive. Unlike the M@ZnO, in the lower energy region (< 3.2 eV), $(ZnO)_{48}$ does not have a major peak and is quite smooth but in the higher energy region (> 10 eV), all of the tendency of curves become consistent with each other. In addition, as the DOS presented in Fig. 4(d), M@ZnO shows a typical half-metallic behavior from

**10 / 22**

spin majority and minority components, which is in keeping with the result of the real part of dielectric function. In addition, the spin polarization of Fe, Co and Ni is the major contribution for DOS around the Femi level.

Furthermore, the imaginary part of dielectric function shows that the curve of $(ZnO)_{48}$ has no distinct peak while M@ZnO appears to have a larger peak at around 0.85-1.47 eV, which is mainly due to the contribution of Co, Fe, Ni atoms in the core. It indicates that there is an evident absorptive action in the infrared region and the margin of visible light, especially in the case of $Fe_{15}@(ZnO)_{42}$, whose peak of absorption is closer to the visible light region. Finally, due to the interaction between O atom in shell and metal atom in core, the peak at 9.68 eV of $(ZnO)_{48}$ vanishes and the curve decreases to zero rapidly.

## 3. Conclusions

The structural, magnetic and optical properties of $TM_n@(ZnO)_{42}$ (TM = Fe, Co and Ni) core-shell nanostructures are studied by the First-principles calculations. Our results indicate that $Fe_{13}@(ZnO)_{42}$, $Co_{15}@(ZnO)_{42}$ and $Ni_{15}@(ZnO)_{42}$ core-shell nanostructure are the most stable configurations. Compared with $(ZnO)_{48}$ value, the Zn-O bonding of M@ZnO is stronger due to the interaction of TM-O. The special magnetism mainly effect by O atoms and TM atoms, which can be attributed to the strong TM-O hybridization and charge transfer. It is also found that this strong interaction induces some magnetic moment inversion for $Fe_{13}@(ZnO)_{42}$. Furthermore, the optical properties of M@ZnO are systematically investigated based on absorption coefficient.

Compared with the absorption spectrum of the (ZnO)$_{48}$, we find that an obvious red shift has occurred, and it is in accordance with the behavior of the calculated electronic structure.

## 4. Methods

All calculations in this paper are performed in the VASP codes [23, 24] based on density functional theory (DFT) [25, 26] within the projector augmented wave (PAW) [27]. The exchange and correlation potential is treated with the generalized gradient approximation (GGA) methods as described by Perdew–Burke–Ernzerhof (PBE) [28]. The electron wave functions are expanded in plane wave with a cutoff energy of 480 eV. All atoms are fully relaxed and the convergence tolerance for energy and maximum force are set to $1.0 \times 10^{-5}$ eV and $-5 \times 10^{-3}$ eV/Å. For k-point sampling, we use a single Γ point for the geometry optimizations in the first Brillouin zone. Spin-polarization is taken into account in this work. In the calculations, the free TM$_n$@(ZnO)$_{42}$ is located in a rectangular supercell with a size of 30×30×30 Å$^3$. The interaction between periodic images could be neglected on this size.

## Acknowledgements

We thank Xingyun Zhu and Qihang Zhang for helpful discussions.

## Author Contribution

Y. W. Hu, C. T. Ji and Q. Liu performed the DFT calculation. Y. W. Hu, X. X. Wang and J. R. Huo wrote the manuscript. Y. P. Song gave instruction on this work.

## Additional information

**Competing Interest:** The author declare that they have no competing interests

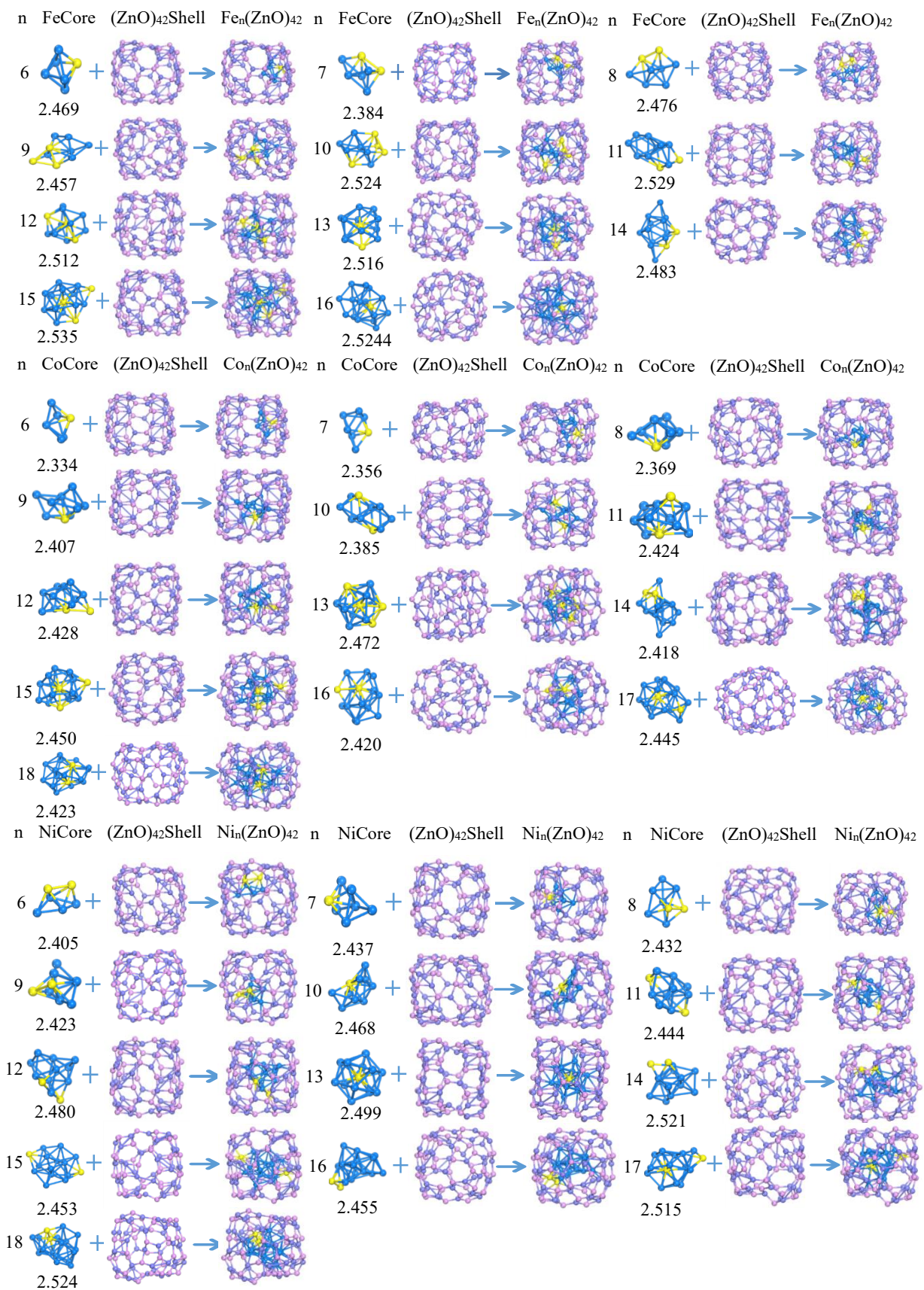

Fig. 1 The optimized geometries of $TM_n@(ZnO)_{42}$ core-shell nanostructure. The pink, purple and blue balls show the positions of O, Zn and TM atoms, respectively. The small or abnormal magnetic moment of TM atoms are shown by yellow balls. The numbers below the inner core configurations indicate the average bond lengths (Å) within 3.00 Å (see Supporting Information I and supporting information III for enlarged picture of core-shell structures).

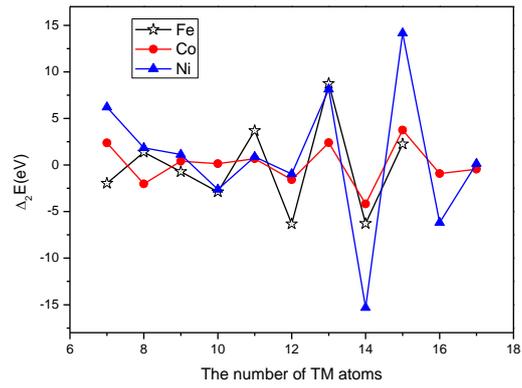

Fig. 2 The second-order differences of total energies $\Delta_2E_n$ of $TM_n@(ZnO)_{42}$ nanostructure. It is noted that the largest $\Delta_2E_n$ are found at n= 13, 15 and 15 for TM = Fe, Co and Ni in $TM_n@(ZnO)_{42}$ core-shell structures, respectively, indicating that $Fe_{13}@(ZnO)_{42}$, $Co_{15}@(ZnO)_{42}$ and $Ni_{15}@(ZnO)_{42}$ are the most stable structure.

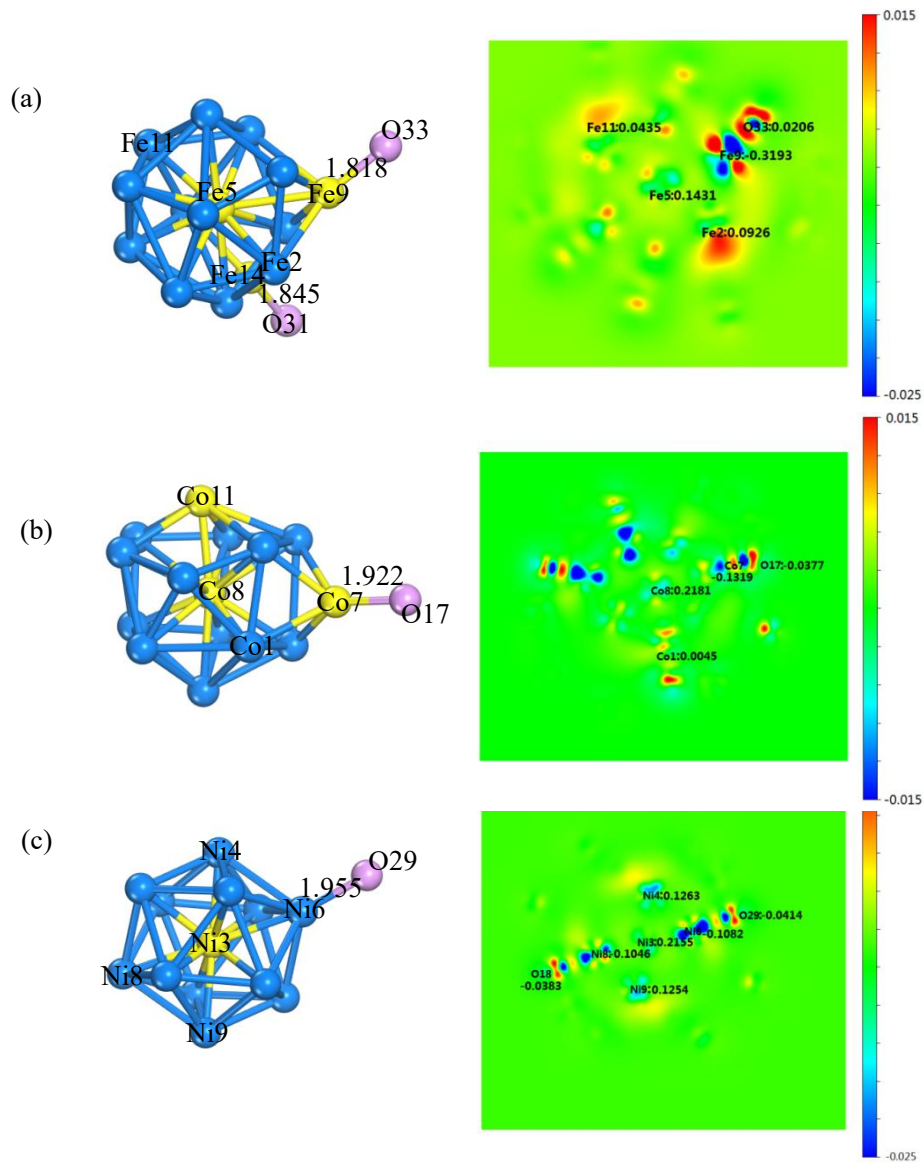

Fig. 3. Plot of the 2D electron density difference and geometry configuration for (a)$Fe_{15}@(ZnO)_{42}$; (b)$Co_{15}@(ZnO)_{42}$ and (c)$Ni_{13}@(ZnO)_{42}$. The atom numbers which overlap with corresponding atom and the bond length (Å) are depicted on geometry configuration

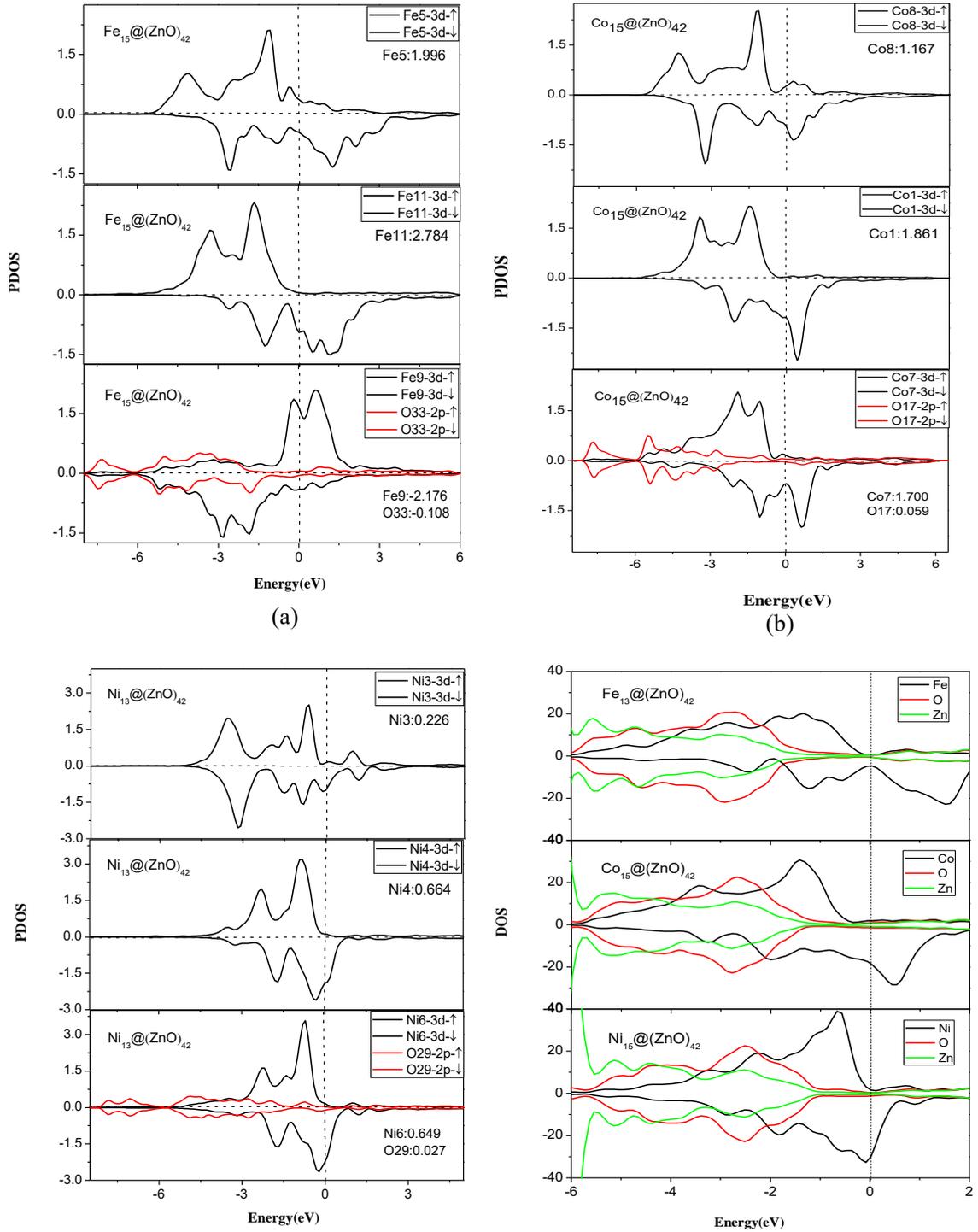

Fig. 4 The partial DOS of (a)Fe$_{15}$@(ZnO)$_{42}$, (b)Co$_{15}$@(ZnO)$_{42}$ and (c)Ni$_{13}$@(ZnO)$_{42}$; and the total DOS of (d)M@ZnO ; The dotted lines refer to the Fermi level. The unit is electrons/eV. The magnetic moment of some corresponding atom is also depicted inside each DOS and the unit is μB/atom.

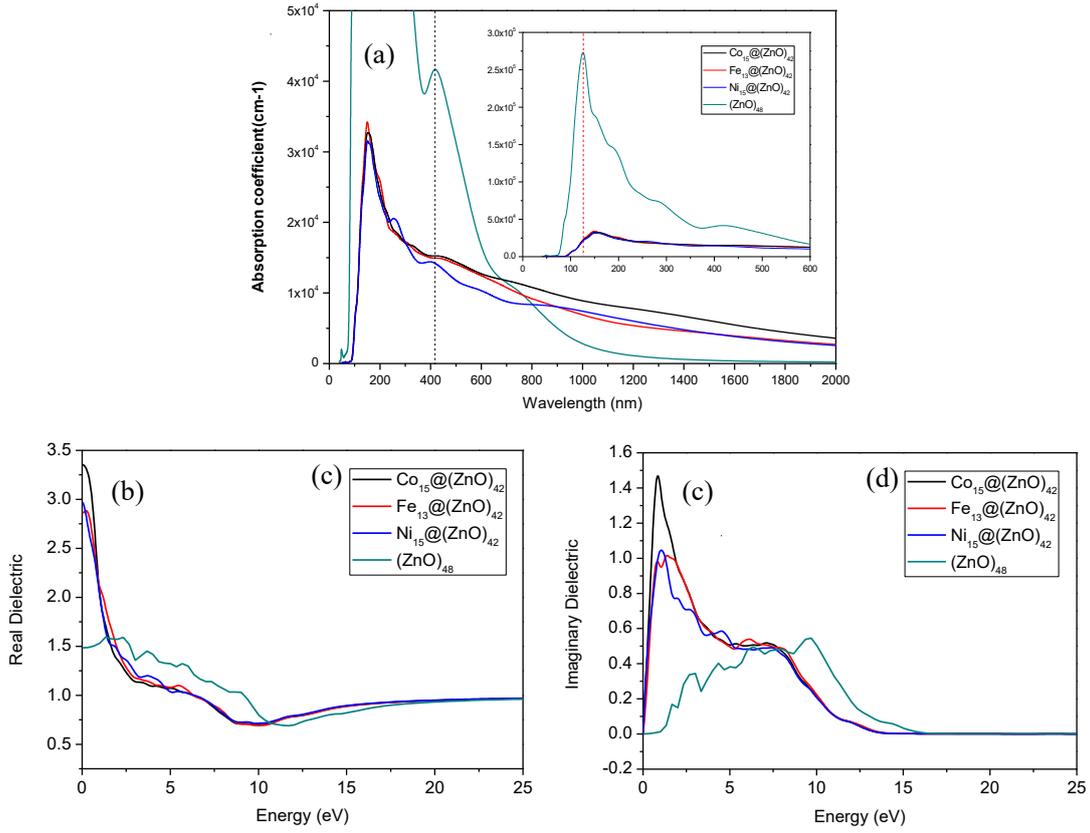

Fig. 5 The calculated optical absorption (a), and dielectric function: real part (b) and imaginary part (c) of M@ZnO (M=Fe,Co,Ni) core-shell structure and (ZnO)$_{48}$.

Table 1 The bond length and bond angle of both the (ZnO)$_{48}$ and M@ZnO core-shell nanostructures, and the values in parenthesis are from the calculations of literature [29].

|  | (ZnO)$_{48}$ | Fe$_{13}$@(ZnO)$_{42}$ | Co$_{15}$@(ZnO)$_{42}$ | Ni$_{15}$@(ZnO)$_{42}$ |
|---|---|---|---|---|
| Zn-O bond length (Å) | 1.875-2.201 (1.842-2.206) | 1.869-2.137 | 1.852-2.177 | 1.859-2.184 |
| O-Zn-O bond angle (°) | 92.3-174.3 (90.2-172.2) | 89.612-161.336 | 88.257-157.0524 | 88.857-157.152 |

Table 2 Atoms magnetic moments (μB) for Fe$_n$@ZnO$_{42}$

|       | n=6   | n=7   | n=8   | n=9    | n=10   | n=11  | n=12   | n=13  | n=14   | n=15   | n=16  |
|-------|-------|-------|-------|--------|--------|-------|--------|-------|--------|--------|-------|
| Fe1   | 2.800 | 2.674 | 2.717 | 2.745  | 2.919  | 2.713 | 2.850  | 2.758 | 2.897  | 2.676  | 2.668 |
| Fe2   | 2.772 | 2.707 | 2.763 | 2.532  | 2.791  | 2.542 | 2.559  | 2.619 | 2.712  | 2.603  | 2.317 |
| Fe3   | 2.631 | 2.750 | 2.717 | 2.832  | **-0.336** | 2.626 | 1.535  | 2.748 | 2.695  | 2.710  | 2.638 |
| Fe4   | 2.748 | 2.379 | 2.778 | 2.707  | **-2.554** | 2.734 | **-2.506** | 2.815 | 2.707  | 2.619  | 2.841 |
| Fe5   | 2.284 | 2.415 | 2.560 | 2.826  | 2.724  | 2.694 | 2.750  | 2.563 | **-1.027** | 1.996 | 1.767 |
| Fe6   | 2.529 | 2.590 | 2.536 | 2.718  | **-0.636** | 2.731 | 2.862  | 2.813 | 2.814  | 2.616  | 2.458 |
| Fe7   |       | 2.830 | 2.674 | 2.599  | 0.335  | 2.678 | 2.898  | 1.537 | 2.863  | 2.772  | 2.355 |
| Fe8   |       |       | 2.827 | 2.789  | 2.149  | 2.653 | 2.770  | 2.665 | 2.761  | 2.332  | 2.366 |
| Fe9   |       |       |       | **-0.023** | 2.468 | 2.782 | 2.652  | 2.786 | 2.599  | **-2.176** | 2.508 |
| Fe10  |       |       |       |        | 2.780  | 2.545 | 2.801  | 2.459 | 2.800  | 2.778  | 2.996 |
| Fe11  |       |       |       |        |        | 2.693 | 2.494  | 2.770 | 2.637  | 2.784  | 2.807 |
| Fe12  |       |       |       |        |        |       | **-2.522** | 2.816 | 2.473  | 2.862  | 2.514 |
| Fe13  |       |       |       |        |        |       |        | 2.544 | 2.789  | 2.340  | 2.571 |
| Fe14  |       |       |       |        |        |       |        |       | 2.831  | **-0.767** | 2.568 |
| Fe15  |       |       |       |        |        |       |        |       |        | 2.577  | 2.699 |
| Fe16  |       |       |       |        |        |       |        |       |        |        | 2.742 |

Table 3 Atoms magnetic moments (μB) for TM$_n$@ZnO$_{42}$ (M=Co,Ni)

|     | Co$_n$@ZnO$_{42}$ | | | | Ni$_n$@ZnO$_{42}$ | | | | | |
|-----|-------|-------|-------|-------|-------|-------|-------|-------|-------|-------|
|     | n=12  | n=13  | n=14  | n=15  | n=8   | n=13  | n=14  | n=15  | n=16  | n=17  |
| M1  | 1.631 | 1.879 | 1.808 | 1.861 | 0.454 | 0.750 | 0.527 | 0.563 | 0.541 | 0.612 |
| M2  | 1.737 | 1.811 | 1.816 | 1.704 | **0.277** | 0.742 | 0.635 | 0.507 | 0.618 | 0.532 |
| M3  | 1.768 | **0.638** | 1.991 | 1.852 | 0.561 | **0.226** | 0.690 | 0.713 | 0.537 | **0.307** |
| M4  | 1.797 | 1.772 | 1.802 | 1.745 | 0.376 | 0.664 | 0.536 | 0.439 | 0.529 | 0.655 |
| M5  | 1.833 | 1.803 | 1.825 | 1.790 | 0.381 | 0.663 | 0.609 | 0.511 | **0.281** | 0.683 |
| M6  | 1.801 | 1.800 | 1.889 | 1.792 | 0.485 | 0.649 | 0.593 | 0.511 | 0.582 | 0.677 |
| M7  | 1.804 | 1.753 | 1.805 | 1.700 | 0.566 | 0.664 | **0.498** | **0.381** | 0.638 | 0.624 |
| M8  | 1.722 | 1.803 | 1.899 | **1.167** | **0.205** | 0.651 | 0.639 | 0.557 | 0.482 | 0.522 |
| M9  | 1.746 | 1.802 | 1.790 | 1.802 |       | 0.663 | **0.458** | 0.580 | 0.504 | 0.548 |
| M10 | 1.674 | 1.770 | 1.963 | 1.816 |       | 0.655 | 0.730 | 0.574 | 0.455 | **0.280** |
| M11 | 1.658 | 1.810 | 1.792 | 1.701 |       | 0.741 | 0.620 | 0.517 | 0.652 | 0.544 |
| M12 | 1.830 | 1.753 | 1.828 | 1.761 |       | 0.655 | 0.606 | 0.618 | **0.380** | 0.543 |

|  | | | | | | | | | |
|---|---|---|---|---|---|---|---|---|---|
| M13 | 1.878 | 1.843 | 1.798 |  | 0.750 | 0.705 | **0.423** | 0.703 | 0.391 |
| M14 |  | 1.859 | 1.785 |  |  | 0.523 | 0.594 | 0.535 | 0.495 |
| M15 |  |  | 1.800 |  |  |  | 0.623 | 0.577 | 0.318 |
| M16 |  |  |  |  |  |  |  | 0.678 | 0.385 |
| M17 |  |  |  |  |  |  |  |  | 0.537 |

Table 4 TM-O distance for $TM_n$@$ZnO_{42}$ (TM = Fe,Co,Ni)

|  | $Fe_{15}$@$ZnO_{42}$ | | $Co_{15}$@$ZnO_{42}$ | | $Ni_{13}$@$ZnO_{42}$ | |
|---|---|---|---|---|---|---|
|  | magnetic moments (μB) | Fe-O distance | magnetic moments(μB) | Co-O distance | magnetic moments (μB) | Ni-O distance |
| M1 | 2.676 | 1.997 | 1.861 | **4.021** | 0.750 | 2.04 |
| M2 | 2.603 | **3.219** | 1.704 | 1.969 | 0.742 | 1.975 |
| M3 | 2.710 | **3.086** | 1.852 | 1.916 | **0.226** | – |
| M4 | 2.619 | 1.915 | 1.745 | 1.945 | 0.664 | **3.364** |
| M5 | 1.996 | – | 1.790 | 2.009 | 0.663 | 1.964 |
| M6 | 2.616 | 1.975 | 1.792 | 1.973 | 0.649 | 1.955 |
| M7 | 2.772 | 1.986 | 1.700 | 1.922 | 0.664 | 1.963 |
| M8 | 2.332 | 2.91 | **1.167** | – | 0.651 | 1.955 |
| M9 | **-2.176** | 1.818 | 1.802 | 1.99 | 0.663 | **3.824** |
| M10 | 2.778 | 1.973 | 1.816 | **3.670** | 0.655 | 2.14 |
| M11 | 2.784 | **3.181** | 1.701 | **3.442** | 0.741 | 1.974 |
| M12 | 2.862 | 1.995 | 1.761 | 1.936 | 0.655 | 2.139 |
| M13 | 2.340 | **3.650** | 1.798 | **3.327** | 0.750 | 2.039 |
| M14 | **-0.767** | 1.845 | 1.785 | 2.002 |  |  |
| M15 | 2.577 | 1.965 | 1.800 | 1.961 |  |  |